\documentclass[aps,pre,amsmath,amssymb,groupedaddress]{revtex4}
\usepackage{graphicx}
\usepackage[utf8]{inputenc}

\usepackage{epsf}
\newcommand{\intlx}{\int\limits_{-\infty}^{\infty}}
\newcommand{\intlo}{\int\limits_{-\infty}^{0}}
\newcommand{\intly}{\int\limits_{0}^{y}}
\renewcommand{\baselinestretch}{1.2}
\date{\today}
\begin{document}%\large
\title{Cauchy Problem for for some high order generalization of Korteweg - de Vries equation}
\author{Z A Sobirov, $|\overline{\underline{\mbox{S Abdinazarov}}}|$}
\affiliation{Mechanics and Mathematics Faculty of National
University of Uzbekistan\\ e-mail: {\it sobirovzar@yahoo.com}}
\begin{abstract}
In this work we study Cauchy problem for a high-order differential
equation $\frac{\partial u(y,x)}{\partial y}+P\left(\frac{\partial
}{\partial x}\right)u(y,x)=\gamma\frac{\partial}{\partial
x}(u^2(y,x))+F(y,x)$. We prove that the problem is well-posed both
for linear ($\gamma =0$) and nonlinear  equations on the class of
rapidly decaying Schwartz functions. Furthermore, for the case
when the initial condition is given on $L_2(\mathbf{R}^1)$ we
prove the existence of the
 unique solution  on the space $L_{\infty}(0,y_0; L_2(\mathbf{R}^1))\bigcap L_2(0,y_0; H^{n-1}(\mathbf{R}^1))\bigcap  L_2(0,y_0;H^{n}(-r, r))$,
 where $r$ is an arbitrary positive number. It is also shown that the solution continuously depends on the initial conditions.
\end{abstract}
 \maketitle


\section{Introduction}

Korteweg - de Vries (KdV) equation is a nonlinear differential
equation that has important application in different areas of
physics(e.g., acoustics, hydrodynamics, optics etc). Therefore
studying its properties is of fundamental and practical
importance. Especially, the problem of soliton transport in
non-uniform media causes special importance of KdV equation.
Different practical applications of this equation and the
properties of its solutions can be found in theRefs. [1], [2], (
see also [7] and references therein).

In this work we treat Cauchy problem for a generalization of
Korteweg - de Vries  equation. Namely, we consider  third order
derivative in the KdV with odd order differential operator with
constant coefficient.

This paper is organized as follows  In the next section we prove
solvability of the Cauchy problem in the Schwartz class of rapidly
decreasing functions. In particular,  subsections A and B deal
with the deal with the linear counterpart,  for which using
Fourier transformation, we show that under certain (necessary and
sufficient) condition for the coefficients, the problem has a
unique solution in the Schwartz class. The  subsection B presents
also some estimates for the Green function.

In subsection I.C  we treat the nonlinear equation by obtaining
countably many set of a-priory estimates implying convergence of
the iteration procedure with respect to nonlinear term in some
interval $(0; y_1);\ y_1 > 0$. We also prove that this solution
can be continued to an arbitrary interval $(0; y_0);\ y_0 > y_1$.
The case of KdV equation ($n=1$) one can find in
\cite{Yakupov1975}.

In the section III we consider Cauchy problem with the initial
function in L2(R1) We explore the case of odd and high than third
order equation. In the subsection IIIA using Green function method
we prove the existence of a weak solution for linear equation.
Also, we obtain some a-priory estimates that will be used in the
further analysis and show continuity of the obtained solution. The
next subsections  present the proofs for the existence of weak
solution for non linear equation and few a-priory estimates.
Finally, using the Green function we prove linear dependence of
the weak solution on initial data.

\section{Solvability of Cauchy problem in the Schwartz class of rapidly decreasing functions. }
\subsection{Well-posedness of linear equation.}\label{1A}
 We consider Cauchy problem for the
equation
\begin{align}
Lu\equiv (-1)^n\frac{\partial u}{\partial
y}+\frac{\partial^{2n+1}}{\partial
x^{2n+1}}+\sum_{k=0}^{2n-1}b_k\frac{\partial^ku}{\partial
x^k}=f(y,x), \label{MainEq}
\end{align}
\begin{align}
u(y,x)|_{y=0}=u_0(y), \label{IC} \end{align} in the half string
$D_{y_0}=\{(y,x):-\infty<x<\infty, 0<y<y_0\}$ where $b_k$ are
constants and $y_0=const>0$.

First we consider the case $f(y,x)\in C^1([0,y_0];
S(\mathbf{R}^1))$, $u_0\in S(\mathbf{R}^1)$.

\textbf{Theorem 1.} {\it Let $\sum_{k=0}^{n-1}
(-1)^{n+k}b_{2k}\lambda^{2k}\geq 0$ for large enough values of
$\lambda
>0$. Then there is unique solution of the Cauchy problem
(\ref{MainEq}), (\ref{IC}) in $C^1([0,y_0]; S(\mathbf{R}^1))$.}

\textbf{Proof.} Using Fourier transform we have
\begin{align}
\frac{d}{dy}\tilde u(y,\lambda)=P(-i\lambda)\tilde u (y,\lambda)
+\tilde f(y,\lambda), \ \ \tilde u(0,\lambda)=\tilde u_0(\lambda),
\label{MEFT}
\end{align}
where $\tilde u, \tilde f$ and  $\tilde u_0$ are Fourier image of
the functions $u, f$ and $u_0$, respectively,
$$P(\lambda)=(-1)^{n+1}\left[\lambda^{2n+1}+\sum_{k=0}^{2n-1}b_k\lambda^k\right]. $$

The unique solution of the problem (\ref{MEFT}) is
\begin{align}
\tilde u(y,\lambda)& = \tilde u_0(\lambda) \exp
(P(-i\lambda)y)+\nonumber\\
&+\int\limits_0^y\tilde f(\eta, \lambda)\cdot
\exp(P(-i\lambda)(y-\eta))d\eta.
 \label{sMEFT} \end{align}

According to the conditions of theorem $\exp (P(-i\lambda)y)$ is
bounded for $y>0$ and its derivatives with respect to $\lambda$
have at most polynomial growth at $|\lambda |\to +\infty$.
Consequently $\tilde u(y,\lambda)\in S(\mathbf{R}^1)$ for any
$y\geq 0$. According to properties of direct and Fourier
transforms we conclude that Cauchy problem (\ref{MainEq}),
(\ref{IC}) has unique solution $u(y,\lambda)\in C^1([0,y_0];
S(\mathbf{R}^1))$.

\subsection{Fundamental solution. Green function for Cauchy problem}

It is known that the solution of model equation
\begin{align}
(-1)^n\frac{\partial u(y,x)}{\partial y}+
\frac{\partial^{2n+1}u(y,x)}{\partial x^{2n+1}}=g(y,x)
\label{modEq}
\end{align}
which satisfies initial condition $u(0,x)=u_0(x)$ given by
\cite{Djur1991}
\begin{align}
&u(y,x)=\int\limits_{-\infty}^{\infty}U(y, x-\xi)u_0(\xi)d\xi+\nonumber\\
& +(-1)^n\int\limits_0^yd\eta
\int\limits_{-\infty}^{\infty}U(y-\eta, x-\xi)g( \eta, \xi)d\xi,
\label{MEsol}
\end{align}
where
\begin{align}
U(y,x)=\pi^{-1} y^{-1/(2n+1)} {\rm
Ain}\left(xy^{-1/(2n+1)}\right), \label{GFME}
\end{align}
is a fundamental solution and
\begin{align}
{\rm Ain}(x)=\int_0^{\infty}cos(\lambda^{2n+1}-\lambda x)d\lambda
\label{ain}
\end{align}
is a Airy function which satisfies the following ordinary
differential equation
\begin{align}
\left(\frac{d^{2n}}{d x^{2n}}+\frac{(-1)^nx}{2n+1}\right)z(x)=0.
\label{Airyeq}
\end{align}

The fundamental solution satisfies the following estimates
\begin{align}
\left|\frac{\partial^jU(y,x)}{\partial x^j}\right|\leq
c_1\frac{x^{-\frac{2(n-j)-1}{4n}}}{y^{\frac{2j+1}{4n}}}, \ \
j=\overline{0,2n-1},  \label{EstMFS+}
\end{align}
for $x>0$ and
\begin{align}
\left|\frac{\partial^jU(y,x)}{\partial
x^j}\right|\leq\frac{c_2}{y^{j+1}{2n}}\exp\left(-c\frac{(-x)^{\frac{2n+1}{2n}}}{y^{\frac{1}{2n}}}\right),
\ \ j=\overline{0,2n-1}, \label{EstMFS-}
\end{align}
for $x<0$, $c, c_1, c_2=const>0$.
\begin{align}
u(y,x)=\int\limits_{-\infty}^\infty U(y,
x-\xi)u_0(\xi)d\xi+\nonumber\\
+(-1)^n\int\limits_0^yd\eta\int\limits_{-\infty}^{\infty}
U(y-\eta, x-\xi)f(\eta\xi)d\xi + \mathcal{J}u(y,x), \label{intEq}
\end{align}
where
\begin{align}
\mathcal{J}u(y,x)= (-1)^n
\int\limits_0^yd\eta\int\limits_{-\infty}^{\infty}\sum_{k=0}^{2n-1}\frac{\partial^k
U}{\partial x^k}(y-\eta, x-\xi)u(\eta, \xi)d\xi. \label{Jey}
\end{align}

According to the theorem 1 integral equation (\ref{intEq}) has
unique solution. Therefore the solution of Cauchy problem can be
presented by the following form
\begin{align}
u(y,x)=\int\limits_{-\infty}^\infty G(y,
x-\xi)u_0(\xi)d\xi+\nonumber\\
+(-1)^n\int\limits_0^yd\eta\int\limits_{-\infty}^{\infty}
G(y-\eta, x-\xi)f(\eta\xi)d\xi,
 \label{SolOfCP} \end{align}
where
\begin{align}
G(y-\eta, x-\xi)= &\int\limits_0^y\int\limits_{-\infty}^{\infty}
R(y-t, x-z)U(t-\eta, z-\xi)dzdt+\nonumber\\
&+U(y-\eta, x-\xi),\nonumber
\end{align}
$R(y-\eta, x-\xi)$ is a resolvent of integral operator
$\mathcal{J}$. It is known that resolvent of integral equation
satisfies the same estimates at infinity as kernel of integral
operator. So,
\begin{align}
|R(y, x)|\leq \left\{ \begin{array}{cc}
c_1\frac{x^{\frac{2n-1}{4n}}}{y^{\frac{1}{2n+1}}}, &
\frac{x}{y^{1/(2n+1)}}\to +\infty;\\
\frac{c_2}{y^{\frac{2n}{2n+1}}}\exp\left(-c\frac{(-x)^{\frac{2n+1}{2n}}}{y^{\frac{1}{2n}}}\right),&
\frac{x}{y^{1/(2n+1)}}\to -\infty
\end{array}\right.
\label{Res_Est}
\end{align}

\textbf{ Proposition. } {\it The function $G$ satisfies the
following estimates
\begin{align}
\left|\frac{\partial^k}{\partial
x^k}G(y,x)\right|\leq\left\{\begin{array}{cc}
c\cdot\frac{x^{1+\frac{k}{2n}}}{y^{\frac{2k+1}{4n}}} & \mbox{ for
}
x>0;\\
\frac{const}{y^{\frac{k+1}{2n+1}}}\exp\left(-c'\frac{(-x)^{\frac{2n+1}{2n}}}{y^\frac{1}{2n}}\right)
&\mbox{ for } x<0,\end{array}\right.
 \label{GFEst} \end{align}
where $k=0;1$.}

\textbf{Proof.} We put
\begin{align}
I(y,x)=\int\limits_0^y dt \int\limits_{-\infty}^\infty R(y-t,
x-z)U(z,t)dz. \label{IntI}
\end{align}

\textit{(a)} Let $x>0$. Then rewrite the integral
$\frac{\partial^k}{\partial x^k}I(y,x)$ in the form
\begin{align}
\frac{\partial^k}{\partial x^k}I(y,x)=\int\limits_0^y dt
\int\limits_{-\infty}^\infty R(y-t, x-z)\frac{\partial^k}{\partial
x^k}U(t,z)dz=\nonumber\\
\int\limits_0^y dt
\left[\int\limits_{-\infty}^0+\int\limits_0^{x}+\int\limits_x^{+\infty}\right]dz\equiv
I_1^{(k)}+I_2^{(k)}+I_3^{(k)}. \label{IntIx>0}
\end{align}

Using the estimates for the functions $R$ and $U$ , using
substitutions $z=-xz_1$ and
$z_1=y^{\frac{1}{2n+1}}{x^{-1}}z_2^{\frac{2n}{2n+1}}$ also taking
to account the identity
$$\int_0^y (y-t)^\alpha t^\beta dt=y^{\alpha +\beta -1} B(\alpha, \beta)$$
 we have
\begin{align}
\left|I_1^{(k)}\right|\leq
c_3\frac{x^{\frac{2n-1}{4n}}}{y^{\frac{k+1}{2n+1}-\frac{1}{4n}}}
\left[\frac{y^{\frac{1}{2n+1}}}{x}\int\limits_0^{+\infty}z^{-\frac{1}{2n+1}}e^{-cz}dz+
\frac{y^{\frac{1}{2n+1}\left(1+\frac{2n-1}{4n}\right)}}{x^{1+\frac{2n-1}{4n}}}\int\limits_0^{+\infty}z^{\frac{1}{2}}e^{-cz}dz\right]\leq
c_5\frac{x^{\frac{2n-1}{4n}}}{y^{\frac{2k+1}{4n}}}.
 \label{EstI1k} \end{align}

Using straightforward calculation we obtain
\begin{align}
\left|I_2^{(k)}\right|\leq
c_6\int\limits_0^y\int\limits_0^x\frac{(x-z)^{\frac{2n-1}{4n}}}{(y-t)^{\frac{4n-1}{4n}}}\cdot\frac{z^{-\frac{2(n-k)-1}{4n}}}{t^{2k+1}{4n}}dzdt
= c_7\frac{x^{1+\frac{k}{2n}}}{y^{\frac{k}{2n}}}. \label{EstI2k}
\end{align}

The integral $I_3^{(k)}$ can be estimated analogously as
$I_2^{(k)}$
\begin{align}
|I_3^{(k)}|\leq c_{11}
\frac{x^{-\frac{2(n-k)-1}{4n}}}{y^{\frac{2k+1}{4n}-\frac{2}{2n+1}}}\left[1+\frac{x^{\frac{2(n-k)-1}{4n}}}{y^{\frac{2(n-k)-1}{4n(2n+1)}}}\right].
\label{EstI3k}
\end{align}

From the estimates (\ref{EstI1k}) -- (\ref{EstI3k}) we obtain
estimation for $\frac{\partial^k}{\partial x^k}G(y,x)$ at $x>0$.
\textit{(b)} Now we consider the case $x<0$. We rewrite the
integral $\frac{\partial^k}{\partial x^k}I(y,x)$ in the form
\begin{align}
\frac{\partial^k}{\partial x^k}I(y,x)=\int\limits_0^y dt
\int\limits_{-\infty}^\infty R(y-t, x-z)\frac{\partial^k}{\partial
x^k}U(t,z)dz=\nonumber\\
\int\limits_0^y dt
\left[\int\limits_{-\infty}^x+\int\limits_x^0+\int\limits_0^{+\infty}\right]dz\equiv
I_4^{(k)}+I_5^{(k)}+I_6^{(k)}. \label{IntIx<0}
\end{align}

Taking to account estimates for the functions $U$ and $R$ and
using $\theta^b\exp(-a\theta)\leq const$ for $\theta >0, a>0$ we
have

\begin{align}
|I_4^{(k)}|\leq
\frac{c_{13}}{y^{\frac{k+1}{2n+1}-\frac{1}{4n}-\frac{2n-1}{4n(2n+1)}}}&\int_{-\infty}^x\exp\left(-(c-\varepsilon)
\frac{(-z)^{\frac{2n+1}{2n}}}{y^{\frac{1}{2n}}}\right)dz=\nonumber\\
&\left\{z=z_1^{\frac{2n}{2n+1}}y^{\frac{1}{2n+1}}\right\}=\nonumber\\
\frac{2n}{2n+1}\frac{c_{13}}{y^{\frac{k}{2n+1}-\frac{1}{4n}-\frac{2n-1}{4n(2n+1)}}}&\int_{-\infty}^{\frac{(-x)^{(2n+1)/(2n)}}{y^{1/(2n)}}}
(-z_1)^{-\frac{1}{2n+1}}\exp(c_0z_1)dz_1\nonumber\\
\leq\frac{c_{14}}{y^{\frac{k}{2n+1}-\frac{1}{4n}-\frac{2n-1}{4n(2n+1)}}}&\int_{-\infty}^{\frac{(-x)^{(2n+1)/(2n)}}{y^{1/(2n)}}}\exp((c_0-\varepsilon)z_1)dz_1
\nonumber\\
\leq\frac{c_{14}}{y^{\frac{k}{2n+1}-\frac{1}{4n}-\frac{2n-1}{4n(2n+1)}}}&\exp\left(\frac{(-x)^{(2n+1)/(2n)}}{y^{1/(2n)}}
\right).
 \label{EstI4k}
\end{align}

Now we estimate $|I_5^{(k)}|$.
\begin{align}
|I_k^{(5)}\leq
c_{15}\int\limits_0^y\frac{dt}{(y-t)^{\frac{2n}{2n+1}}
t^{\frac{k+1}{2n+1}}}&\int_x^0
\exp\left(-c\frac{(z-x)^{\frac{2n+1}{2n}}}{(y-t)^{\frac{1}{2n}}}-
c\frac{(-z)^{\frac{2n+1}{2n}}}{t^{\frac{1}{2n}}}
\right)\nonumber\\
\leq
c_{16}\frac{-x}{y^{\frac{k}{2n+1}}}&\cdot\exp\left(-c'\frac{(-x)^{\frac{2n}{2n+1}}}{y^{\frac{1}{2n}}}\right).
\label{EstI5k}
\end{align}
Here we use inequalities $y-t<y, \ t<y$ and
$a^{\alpha}+b^{\alpha}\geq \left(\frac{a+b}{2}\right)^{\alpha}$
for $a>0, b>0, \alpha >0$.

Analogously as in the case of integral $I_4^{(k)}$ we get
\begin{align}
|I_6^{(k)}|\leq
\frac{c_{19}}{y^{\frac{2k+1}{4n}-\frac{2}{2n+1}-\frac{2(n-k)-1}{4n(2n+1)}}}\exp\left(-c_0\frac{(-x)^{\frac{2n+1}{2n}}}{y^\frac{1}{2n}}\right).
\label{EstI6k}
\end{align}

From the estimates (\ref{EstI4k}) -- (\ref{EstI6k}) we get
estimates for $\frac{\partial^k}{\partial x^k}G(y,x)$ at $x<0$.

\subsection{Cauchy problem for nonlinear equation in the class of rapidly decaying functions}
In this section we investigate Cauchy problem for nonlinear
equation
\begin{align}
Lu(y,x)=\gamma\frac{\partial}{\partial x} \left(u^2(y,x)\right)
+F(y,x), \ \ x\in \mathbf{R}^1 , y>0\label{NLEq}
\end{align}
where $F(y,x)\in C^1([0,y_0]; S(\mathbf{R}^1))$, with initial
condition $u(0,x)=u_0(x)\in S(\mathbf{R}^1)$.

\textbf{Theorem 2.} {\it (Uniqueness of solution) Let
$(-1)^{n+k}b_{2k}\geq 0$ ($k=overline{0,n-1}$). Then the Cauchy
problem (\ref{NLEq}), (\ref{IC}) has at most one solution in
$C^1([0,y_0], S(\mathbf{R}^1))$ for any $y_0>0$.}

\textbf{Proof.} Let suppose that $u_1(y,x)$ and $u_2(y,x)$ are two
different solutions of the Cauchy problem for equation
(\ref{NLEq}). Then the function $u(y,x)=u_1(y,x)-u_2(y,x)$
satisfies the equality
\begin{align}
Lu=2\gamma u \frac{\partial u_1}{\partial x} -2\gamma
\frac{\partial u_2}{\partial x}\label{EqForDiff}
\end{align}
and the initial condition $u(0,x)=0$.

Multiplying both sides of equality (\ref{EqForDiff}) by $(-1)^n
u(y,x)$ and integrating with respect to variable $x$ from
$-\infty$ to $+\infty$ we get
\begin{align}
&\frac{1}{2}\frac{\partial}{\partial
y}\int\limits_{-\infty}^{\infty} u^2dx
+\sum_{k=0}^{n-1}b_{2k}\int\limits_{-\infty}^{\infty}\left(\frac{\partial^k
u}{\partial x^k}\right)dx=\nonumber\\
&=2\gamma\int\limits_{-\infty}^{\infty}u^2\left[2\frac{\partial
u_1}{\partial x}- \frac{\partial u_2}{\partial
x}\right] dx\leq\nonumber\\
\leq&\left(2{ \sup\limits_{x}}\left|\frac{\partial u_1}{\partial
x}\right|+{ \sup\limits_x}\left|\frac{\partial u_2}{\partial
x}\right|\right)\int\limits_{-\infty}^{\infty}u^2 dx\leq A_6
\int\limits_{-\infty}^{\infty}u^2 dx. \label{EE0}
\end{align}

According to conditions of theorem we have
$$\frac{\partial}{\partial y}\int\limits_{-\infty}^{\infty}u^2(y,x)dx\leq A_6 \int\limits_{-\infty}^{\infty}u^2(y,x)dx.$$
It follows
$$e^{-A_6y}\int\limits_{-\infty}^{\infty}u^2(y,x)dx\leq \int\limits_{-\infty}^{\infty}u^2(0,x)dx=0.$$
or $u(y,x)\equiv 0$. This proves the theorem.

It is known that convergence in $C^1([0,y_0], S(\mathbf{R}^1))$
are defined by countable set of semi-norms (see \cite{Shabat1973},
\cite{Yakupov1975})
$$\||u\||_{k,s,j}^2=\sup\limits_{y}\left[\int\limits_{-\infty}^\infty\left|\frac{\partial^{k+j}u(y,x)}{\partial x^k\partial y^j}\right|^2dx+
\int\limits_{-\infty}^\infty(1+x^2)^s\left|\frac{\partial^{j}u(y,x)}{\partial
y^j}\right|^2dx \right],$$ where $k,s$ are nonnegative integers
and $j=0; 1.$

For the further results we need the following

\textbf{Lemma 1 \cite{Yakupov1975}.}{\it Suppose $u\in
S(\mathbf{R}^1)$ and for some $N$ the inequality
$\int_{-\infty}^\infty\left(\frac{\partial^Nu}{\partial
x^N}\right)dx\leq C=const$. Then the following inequality is true
\begin{align}
\int\limits_{-\infty}^\infty x^{2m}& \left(\frac{\partial^k
u}{\partial x^k}\right)^2dx\leq
C_1(k,m)\left(\int\limits_{-\infty}^\infty
x^{2m+2}u^2dx\right)^{\frac{m}{m+1}}+\nonumber\\
&+ C_2(k,m)\left(\int\limits_{-\infty}^\infty
x^{2m+2}u^2dx\right)^{\frac{1}{2^k+1}} \label{L1}
\end{align}
for $2mk\leq N$, where $C_1(m,k), C_2(m,k)$ and $C$ are some
positive constants, $m, k$ and $N$ are natural numbers.}

It is easy to see that from (\ref{L1}) it follows that
\begin{align}
\int\limits_{-\infty}^\infty x^{2m}& \left(\frac{\partial^k
u}{\partial x^k}\right)^2dx\leq
M_{\varepsilon}(m,k)+\varepsilon\int\limits_{-\infty}^{\infty}x^{2m+2}u^2dx,
\label{L1_2}
\end{align}
where $\varepsilon$ and $M_{\varepsilon}(m,k)$ are positive
constants.

\textbf{Theorem 3.} {\it Let $(-1)^{n+k}b_{2k}>0$ ($k=\overline{0,
n-1}$). Then there exist positive $y_1$ which depends on the
coefficients of the equation (\ref{NLEq}) and on quantities
$$\int\limits_{-\infty}^{\infty}\left[u_0^2(x)+\left(\frac{d^2
u_0(x)}{dx^2}\right)^2\right]dx, \ \
\sup\limits_{y}\int\limits_{-\infty}^\infty
\left[F^2(y,x)+\left(\frac{d^2 F(y,x)}{dx^2}\right)^2\right]dx
$$
such that Cauchy problem (\ref{NLEq}), (\ref{IC}) has a solution
in $C^1([0,y_1]; S(\mathbf{R}^1))$.}

\textbf{Proof.} The case $n=1$ are investigated in
\cite{Shabat1973} and \cite{Yakupov1975}. So, we will consider the
case $n>1$. We construct sequence $\{u_m(y,x): m\in\mathbf{N}\}$:

$u_1(y,x)=u_0(x)$, $u_m(y,x)$ ($m\geq 2$) is a solution of the
following problem
\begin{align}
Lu_m(y,x)=\gamma\frac{\partial}{\partial
x}\left(u_{m-1}^2(y,x)\right)+F(y,x), \ \ u_m(0,x)=u_0(x).
\label{SeqSR}
\end{align}

We will show convergence of this sequence in $C^1([0,y_1],
S(\mathbf{R}^1))$.

Multiplying both sides of equation (\ref{SeqSR}) by
$(-1)^n2u_m(y,x)$ and integrating in $\mathbf{R}^1$ we have
\begin{align}
&2\sum_{k=0}^{n-1}(-1)^{n+k}b_{2k}\int\limits_{-\infty}^{\infty}
\left(\frac{\partial^k u_m(y,x)}{\partial
x^k}\right)^2dx+\frac{\partial}{\partial
y}\int\limits_{-\infty}^\infty u_m^2(y,x)dx=\nonumber\\
= (-1)^n&2\gamma\int\limits_{-\infty}^\infty
u_m(y,x)\frac{\partial}{\partial
x}\left(u_{m-1}^2(y,x)\right)dx+(-1)^n2\int\limits_{-\infty}^\infty
u_m(y,x)F(y,x)dx.\label{SeqSR1}
\end{align}

First we consider the first integral on the right hand side of the
equality
$$\int\limits_{-\infty}^\infty
u_m(y,x)\frac{\partial}{\partial
x}\left(u_{m-1}^2(y,x)\right)dx\leq 2
\sup\limits_{x}|u_{m-1}(y,x)|\cdot\int\limits_{-\infty}^\infty
|u_m|\left|\frac{\partial u_{m-1}}{\partial x}\right| dx.
$$

We should estimate $\sup\limits_{x}|u_{m-1}(y,x)|$. Taking to
account relation $\lim_{x\to -\infty} u(y,x)=0$ we have
\begin{align}
\sup\limits_{x}|u_{m-1}(y,x)|\leq
\left(2\int\limits_{-\infty}^{\infty}\left|u_{m-1}\frac{\partial
u_{m-1}}{\partial x }\right|dx\right)^{\frac{1}{2}} \leq
\left(4\int\limits_{-\infty}^{\infty}|u_{m-1}|^2dx\int\limits_{-\infty}^{\infty}\left|\frac{\partial
u_{m-1}}{\partial x }\right|^2dx\right)^{\frac{1}{4}}
\label{sup(m-1)}
\end{align}

Taking to account (\ref{SeqSR1})-(\ref{sup(m-1)}) and using Cauchy
$\varepsilon$ $ab\leq \varepsilon a^2+\frac{1}{4\varepsilon}b^2$,
 Holder inequality $ab\leq \frac{a^p}{p}+\frac{b^q}{q},
\frac{1}{p}+\frac{1}{q}=1$ and inequality
\begin{align}
\int\limits_{-\infty}^\infty (v'(x))^2dx=
-\int\limits_{-\infty}^\infty v(x) v''x)dx\leq
\left(\int\limits_{-\infty}^\infty v^2(x) dx
\int\limits_{-\infty}^\infty (v''(x))^2 dx\right)^{\frac{1}{2}}
\label{v'} \end{align}  we get

\begin{align}
&2\sum_{k=0}^{n-1}(-1)^{n+k}b_{2k}\int\limits_{-\infty}^{\infty}
\left(\frac{\partial^k u_m(y,x)}{\partial
x^k}\right)^2dx+\frac{\partial}{\partial
y}\int\limits_{-\infty}^\infty u_m^2(y,x)dx\nonumber \\
\leq \varepsilon & \int\limits_{-\infty}^\infty u_m^2dx
+C_0(\varepsilon)
\left[\left(\int\limits_{-\infty}^\infty\left(\frac{\partial^2
u_{m-1}}{\partial
x^2}\right)^2\right)^2+\left(\int\limits_{-\infty}^\infty
u_{m-1}^2dx\right)^2\right]+C_1. \label{est_u_m}
\end{align}

Similarly, taking second partial derivative of both sides of
equation with respect to $x$, multiplying both sides of resulting
equation by $2(-1)^n\frac{\partial^2 u_m}{\partial x^2}$ then
integrating in $\mathbf{R}^1$ and using similar operation as above
we get

\begin{align}
&2\sum_{k=0}^{n-1}(-1)^{n+k}b_{2k}\int\limits_{-\infty}^{\infty}
\left(\frac{\partial^{k+2} u_m(y,x)}{\partial
x^k}\right)^2dx+\frac{\partial}{\partial
y}\int\limits_{-\infty}^\infty \left(\frac{\partial^2 u_m(y,x)}{\partial x^2}\right)^2 dx\nonumber \\
\leq \varepsilon  \int\limits_{-\infty}^\infty
\left(\frac{\partial^3 u_m}{\partial x^3}\right)^2dx&+ \varepsilon
 \int\limits_{-\infty}^\infty \left(\frac{\partial^2
u_m}{\partial x^2}\right)^2dx +C_2(\varepsilon)
\left[\left(\int\limits_{-\infty}^\infty\left(\frac{\partial^2
u_{m-1}}{\partial
x^2}\right)^2\right)^2+\left(\int\limits_{-\infty}^\infty
u_{m-1}^2dx\right)^2\right]+C_3 \label{Est_d^2u_m} \end{align}

Finally we have
\begin{align}
\frac{\partial}{\partial
y}\int\limits_{-\infty}^\infty\left[\left(\frac{\partial^2
u_{m}}{\partial x^2}\right)^2+ u_{m}^2\right]dx\leq C_4
\left[\int\limits_{-\infty}^\infty\left[\left(\frac{\partial^2
u_{m-1}}{\partial x^2}\right)^2+ u_{m-1}^2\right]dx\right]^2+C_5,
\label{Est_d^2u_m+u_m}
\end{align}
where $C_4$ is a constant that depends on the coefficients  $b_2,
b_0, \gamma$ of equation (\ref{NLEq}) and
$$C_5=\frac{1}{\varepsilon}\sup\limits_y\int\limits_{-\infty}^\infty \left(F^2(y,x)+\left(\frac{\partial^2 F(y,x)}{\partial x^2}\right)^2\right)dx$$

Let suppose
\begin{align}
\sup\limits_y
\int\limits_{-\infty}^\infty\left[\left(\frac{\partial^2
u_{m-1}}{\partial x^2}\right)^2+ u_{m-1}^2\right]dx\leq 2
\int\limits_{-\infty}^\infty\left[\left(\frac{d^2 u_{0}}{d
x^2}\right)^2+ u_{0}^2\right]dx +C_5.\label{IndStep_1} \end{align}
Then one can easily check that
\begin{align}
\sup\limits_y
\int\limits_{-\infty}^\infty\left[\left(\frac{\partial^2
u_{m}}{\partial x^2}\right)^2+ u_{m}^2\right]dx\leq 2
\int\limits_{-\infty}^\infty\left[\left(\frac{d^2 u_{0}}{d
x^2}\right)^2+ u_{0}^2\right]dx +C_5.\label{IndStep_2} \end{align}
for
$$y\leq y_1=\left(4C_4 \left(\int\limits_{-\infty}^\infty\left[\left(\frac{d^2 u_{0}}{d
x^2}\right)^2+
u_{0}^2\right]dx\right)^2+2C_5^2+C_5\right)^{-1}\cdot
\left(\int\limits_{-\infty}^\infty\left[\left(\frac{d^2 u_{0}}{d
x^2}\right)^2+ u_{0}^2\right]dx+C_5\right).$$

Since the inequality (\ref{IndStep_2})  is true for $m=1$ and
$y_1$ does not depend on $m$ we conclude that it is hold for any
$m\in \mathbf{N}, \ 0\leq y\leq y_1$.

Now we carry out mathematical induction with respect to order of
derivative. Let suppose that
$\sup\limits_y\int\limits_{-\infty}^\infty \left(\frac{\partial^j
u_m}{\partial x^j}\right)^2dx\leq C$ for any $j<l, \ m\in
\mathbf{N}$, where $C$ is constant which is does not dependent on
$m$.

Using similar operation as above and taking to account inequality
(\ref{IndStep_2}) one can easily obtain
$$\frac{\partial }{\partial y}\int\limits_{-\infty}^{\infty} \left(\frac{\partial^l u_m}{\partial x^l}\right)^2dx
\leq M_4 \int\limits_{-\infty}^{\infty} \left(\frac{\partial^l
u_{m-1}}{\partial x^l}\right)^2dx+M_5$$ which implies
$\int\limits_{-\infty}^{\infty} \left(\frac{\partial^l
u_m}{\partial x^l}\right)^2dx\leq const$ for all $0\leq y \leq
y_1$.

Analogously can be shown that $\int\limits_{-\infty}^\infty
x^{2p}u_m^2dx\leq const$. Then according to lemma 1 we have
$$\int\limits_{-\infty}^{\infty}x^{2p}\left(\frac{\partial^ku_m}{\partial x^k}\right)^2dx\leq
\int\limits_{-\infty}^{\infty}x^{2p+2}u_m^2dx +A_4\leq const<
\infty .$$

According to the estimates given above the sequence $\{ u_m\}$
converges in $C^1([0,y_1]; S(\mathbf{R})^1)$. It can be easily
checked that the function $u:=\lim\limits_{n\to +\infty}u_n$ is a
solution of Cauchy problem (\ref{NLEq}), (\ref{IC}).

Now we will show solvability of Cauchy problem in $0<y<y_0$ for
any $y_0 >0$.

\textbf{Theorem 4.} {\it Suppose $(-1)^{n+k} b_{2k}>0$
($k=\overline{0, n-1}$), $u_0(x)\in S(\mathbf{R}^1)$ and
$F(y,x)\in C^1([0,y_0]; S(\mathbf{R}^1))$. Then Cauchy problem
(\ref{NLEq}), (\ref{IC}) has a solution in $C^1([0,y_0];
S(\mathbf{R}^1))$.}

\textbf{Proof.} We show that the solution obtained above can be
continued to the interval $0<y<y_0$ for any $y_0>0$. According the
proof of Theorem 3 it is enough to estimate the norm of $u$ in
$L_{\infty}([0,y_0]; H^3(\mathbf{R}^1))$. Multiplying equation
(\ref{NLEq}) by $(-1)^n2u(x,y)$, integrating in $D_y=(0,y)\times
\mathbf{R^1}$ and  integrating by part we have
$$2\sum_{k=0}^{n-1}(-1)^{n+k}b_{2k} \int\limits_{D_y}\left(\partial_{x}^k u(\eta ,
y)\right)^2dxd\eta +(1-\varepsilon )\int\limits_{-\infty}^{\infty}
u^2(y,x)dx\leq \int\limits_{-\infty}^{\infty} u_0^2(x)dx+
C(\varepsilon , y_0)\int\limits_{D_{y_0}}F^2(y,x)dxdy:= l_1(y_0,
\varepsilon),
$$
where $\partial_x :=\frac{\partial}{\partial x}$. It follows

$$\int\limits_{D_y}\left(\partial_x^k u(\eta ,x)\right)dxd\eta\leq\frac{1}{(-1)^{n+k}b_{2k}}l_1(y_0,
\varepsilon), \ \ \int\limits_{-\infty}^\infty u^2(y,x)dx\leq
l_1(y_0, \varepsilon)$$ $k=\overline{0, n-1}, \ 0<y<y_0$.

Taking derivative $\partial_x$ from both sides of equation
(\ref{NLEq})
\begin{align}
&2\sum_{k=0}^{n-1}(-1)^{n+k}b_{2k}
\int\limits_{D_y}\left(\partial_{x}^{k+1} u(\eta ,
y)\right)^2dxd\eta +(1-\varepsilon )\int\limits_{-\infty}^{\infty}
(\partial_x u(y,x))^2dx\nonumber\\
\leq 4&(-1)^n\gamma\int\limits_{D_y}(\partial_x u(\eta,
x))^3dxd\eta +\int\limits_{-\infty}^{\infty}
(u_0^{\prime})^2(x)dx+ C(\varepsilon ,
y_0)\int\limits_{D_{y_0}}(\partial_x F(y,x))^2dxdy \label{q1}
\end{align}

\begin{align}
\int\limits_{D_y}(\partial u(\eta,
x))^3dxd\eta\leq\int\limits_0^y\sup_{x}|\partial u(\eta
,x)|\int\limits_{-\infty}^{\infty}(\partial u(\eta
,x))^2dxd\eta\leq\nonumber\\
\int\limits_0^y \left(\intlx (\partial_xu(\eta
,x))^2dx\right)^{\frac{5}{4}}\left(\intlx(\partial_x^2(\eta,
x))^2dx\right)^{\frac{1}{4}}\nonumber\\\leq
c_6(\varepsilon)\intly\left(\intlx(\partial_x(\eta
,x))^2dx\right)^{\frac{5}{3}}d\eta+
+\varepsilon\intly\intlx(\partial_x
u(x,\eta))^2dxd\eta\nonumber\\
\leq \varepsilon\sup_{\eta\in (0,y_0)}\intlx(\partial_x u(\eta
,x))^2dx+\varepsilon\intly\intlx(\partial_x^2u(\eta,x))^2dx
+c_8(\varepsilon, y_0). \label{q2}
\end{align}

According inequalities (\ref{q1}) and (\ref{q2}) we have
\begin{align}
\int\limits_{D_y}\left(\partial_x^{k+1} u(\eta
,x)\right)dxd\eta\leq const, \ \ \int\limits_{-\infty}^\infty
(\partial u(y,x))^2 dx\leq const,\ \ k=\overline{0, n-1}, \
0<y<y_0. \label{u_xx}\end{align}

Analogously one can get
$$\sup_{y\in(0,y_0)}\intlx(\partial_x^2u(y,x))^2dx\leq C.$$

That proves the theorem.

It should be noticed that if $F(y,x)\in C^{\infty}([0,y_0];
S(\mathbf{R}^1))$ the Cauchy problem has a solution in
$C^{\infty}([0,y_0]; S(\mathbf{R}^1))$.

\section{Weak solution of the Cauchy problem}

In this section we investigate Cauchy for the equation
\begin{align}
Lu=\gamma\partial_x (u^2) \label{NLEq2}  \end{align} $(y,x)\in
D_{y_0}:=(0,y_0)\times \mathbf{R}^1$, $y_0>0$, with initial
condition
\begin{align}
u(y,x)|_{y=0}=u_0(x)\in L_2(\mathbf{R}^1). \label{IC2} \end{align}

\textbf{Definition.} {\it The function $u(y,x)\in L_2(D_{y_0})$ is
called to be weak solution of Cauchy problem (\ref{NLEq2}),
(\ref{IC2}) if the following conditions are hold

(a) for any function $\varphi (y, x)\in C_0^{\infty}(D_{y_0})$
\begin{align}
{\int\int}_{D_{y_0}}\left(u\cdot L^*\varphi +\gamma
u^2\partial_x\varphi\right)dxdy=0, \label{IntIdent}
\end{align}
where
$$L^*=-\partial^{2n+1}-(-1)^n\partial_y +\sum_{k=0}^{n-1}(-1)^kb_k\partial_x^k;$$

 (b) There exist a set $E\subset (0,y_0), \ {\rm mes} E=0$ such that
 for any $y\in (0,y_0)\setminus E$ the function $u(y,x)$ is
 well-defined a.e. in $\mathbf{R}^1$ and for any function $\omega (x)\in
 C_0^{\infty}(\mathbf{R}^1)$ the following equality
 \begin{align}
\lim_{\begin{array}{c} y\to 0\\ y\in (0,y_0)\setminus
E\end{array}}\intlx u(y,x)\omega(x)dx = \intlx u_0(x)\omega(x)dx
 \label{IntIC}
 \end{align}
is hold.}

Further in this section we will define by $C(\cdot ,\cdot, ...)$
different positive constants than depends on entering parameters.

For any function $v(x)\in L_2(\mathbf{R}^1)$ and $\alpha >0$ we
put

$$\|v\|=\intlx v^2(x)dx, \ \ N_{\alpha}(v)=\intlo |x|^{\alpha}v^2(x)dx.$$

 Let $\psi_0(x)\in C^\infty(\mathbf{R}^1)$ is a non decreasing
 function such that $\psi_0(x)=0$ for $x\leq\frac{1}{2}$,
 $\psi_0(x)=1$ for $x\geq 0$ and it strictly increase in $\frac{1}{2}\leq x\leq
 1$. For $\alpha >0$ we put $\psi_{\alpha}:=x^{\alpha}\cdot
 \psi_0(x)$. It is easy to see that $\psi_\alpha (x)\in
 C^{\infty}(\mathbf{R}^1)$ and $\psi_\alpha^{\prime}\geq 0$.

 First we will consider the case of linear equation ($\gamma =0$).

 \subsection{Well-posedness of Cauchy problem for linear
 equation}\label{2A}

In this subsection we will consider the case $\gamma =0$ i.e. the
case of linear equation.

{\bf Theorem 5.} {\it Let $ (-1)^{n+k}b_{2k}\geq 0 \
(k=\overline{0,n-1})$ and there exist constant $\varepsilon >0$
such that $N_{3+\varepsilon} (u_0(x))<\infty$. Then
\begin{align}
u(y,x)=\intlx G(y, x-\xi) u_0(\xi) d\xi
 \label{LE1} \end{align}
is a unique weak solution of the Cauchy problem and
\begin{align}
{\rm ess}\sup_{0<y<y_0}\|u(y,x\|\leq c\|u_0\|. \label{LE2}
\end{align}
 Furthermore, if for some $\alpha >0$ the quantity $N_\alpha
 (u_0)$is bounded and $(-1)^{n+k}b_{2k}>0 \ (k=\overline{0, n-1})$
 then
 \begin{align}
{\rm ess}\sup_{0<y<y_0}N_\alpha (u(y,x))\leq C(\alpha,
y_0)\left[\|u_0\|+N_\alpha(u_0)\right].
 \label{LE3}
 \end{align}}

\textbf{Proof.} Theorem 1 implies that adjoint Cauchy problem with
initial condition in $S(\mathbf{R}^1)$ has unique solution in
$C^1([0,y_0] ; S(\mathbf{R}^1))$. Then according to
\cite{Gelfand1953} -- \cite{Gelfand1958} we conclude that Cauchy
problem (\ref{NLEq2}), (\ref{IC2}) has unique solution in the
class of function adjoint to $C^1([0,y_0] ; S(\mathbf{R}^1))$
given by (\ref{LE1}).

Now it is enough to establish estimates (\ref{LE2}), (\ref{LE3}).
We put
$$u_0^h(x):=\frac{1}{h}\intlx\lambda\left(\frac{x-\xi}{h}\right)u_0(\xi)d\xi,$$
where $h>0, \lambda(x)\geq 0, \lambda(x)\in C_0^\infty
(\mathbf{R}^1), {\rm supp} \lambda(x)\subseteqq[-1,1], \intlx
\lambda(x)dx =1.$
$$u_{0h}(x):=u_0^h(x)\cdot\psi_0(x+1/h)\cdot\psi_0(1/h-x),\ \ 0<h<1.$$

We will investigate the following problem
\begin{align}
Lw_h(y,x)=0, \ \ w_h|_{y=0}=u_{0h}(x)\in C_0^\infty(\mathbf{R}^1).
\label{CP_h}
\end{align}

The function $w_h(y,x) =\intlx G(y, x-\xi) u_{0h}(\xi)d\xi$ is a
solution of this problem and according to result of previous
section $w_h\in C^{\infty}(0,y_0: S(\mathbf{R}^1))$ and
$\|w_h\|\leq\|u_0\|$.

Multiplying both sides of equation (\ref{CP_h}) by
$w_h(y,x)\psi_\alpha(1-x)$ and integrating in $\mathbf{R}^1$ we
have
\begin{align}
\frac{1}{2}\intlx w_h^2(y,x)\psi_\alpha(1-x)dx
+\frac{2n+1}{2n}\intly\intlx\left(\partial^n
w_h\right)^2\psi_\alpha^\prime(1-x)dxd\eta+\nonumber\\
+ \sum_{k=1}^{n-1}(-1)^{n+k}b_{2k}\intly\intlx (\partial_x^k
w_h)^2\psi_\alpha(1-x)dxd\eta\leq\intlx
u_{0h}^2(x)\psi_\alpha(1-x)dx+\nonumber\\
+\sum_{k=0}^{n-1}\sum_{1\leq m+2k\leq 2n+1} C_{km}\intly\intlx
(\partial_x^k w_h)^2\psi_\alpha^{(m)}(1-x)dxd\eta\label{CPh_ineq}
\end{align}
where $C_{mk}$ are constants that depends on $m$ and the
coefficients of equation.

First we consider the case $0<\alpha\leq 1$. Then
$\left|\psi_\alpha^{(k)}(x)\right|\leq C(\alpha)$ for all
$k=\overline{1, 2n+1}$. Therefore, inequality (\ref{CPh_ineq})
implies

$$\intly\intlx(\partial_x^k w_h)^2\psi_\alpha(1-x)dxd\eta\leq C(\alpha, y_0)\cdot \|u_0\|^2, \ \ k=\overline{1,n-1},$$
$$\intlx w_h^2(y,x)\psi_\alpha(1-x)dx \leq C(\alpha, y_0)\cdot\|u_0\|^2.$$

Let suppose that the inequalities

\begin{align}
\intly\intlx(\partial_x^k w_h)^2\psi_\alpha(1-x)dxd\eta\leq
C(\alpha, y_0)\cdot\left[ \|u_0\|^2+N_\alpha(u_0)\right], \ \
k=\overline{1,n-1}, \label{P1} \end{align}
\begin{align}
\intlx w_h^2(y,x)\psi_\alpha(1-x)dx \leq C(\alpha, y_0)\cdot
\left[ \|u_0\|^2+N_\alpha(u_0)\right]. \label{P2} \end{align}
are
hold for $0<\alpha\leq p$, $p$ is positive integer.

Then for $p<\alpha\leq p+1$ taking to account inequality
$|\psi_\alpha^{(k)}|\leq C(\alpha) \left(1+\psi_{\alpha
-1}(x)\right)$ we get

\begin{align}
\frac{1}{2}\intlx w_h^2(y,x)&\psi_\alpha(1-x)dx
+\frac{2n+1}{2n}\intly\intlx\left(\partial^n
w_h\right)^2\psi_\alpha^\prime(1-x)dxd\eta+\nonumber\\
+ &\sum_{k=1}^{n-1}(-1)^{n+k}b_{2k}\intly\intlx (\partial_x^k
w_h)^2\psi_\alpha(1-x)dxd\eta\leq\nonumber\\
\leq C(\alpha , y_0)\sum_{k=0}^{n-1}&\sum_{1\leq m+2k\leq 2n+1}
\intly\left[\intlx(\partial_x^k w_h)^2dx+\intlx\psi_{\alpha
-1}(1-x)(\partial_x^k w_h)^2dx\right]d\eta+\nonumber\\
&+\intlx u_0^2(x)\psi_\alpha(1-x)dx\leq C(\alpha,
y_0)\left[\|u_0\|^2+N_\alpha(u_0)\right].
 \label{IndStepP_2}
\end{align}
From the last inequality one can conclude that the  inequalities
(\ref{P1}), (\ref{P2}) are true for any positive integer $p$. The
estimates (\ref{LE2}) and (\ref{LE3}) follow from inequalities
$\|w_h\|\leq\|u_0\|$ and (\ref{P2}).

\textbf{Theorem 6.} (Continuity of the solution) {\it Let $
\sum\limits_{k=0}^{n-1}(-1)^{n+k}b_{2k}\lambda^k\geq 0$ for large
enough values of $\lambda$ and there exist constant $\varepsilon
>0$ such that $N_{3+\varepsilon} (u_0(x))<\infty$. Then
the solution $u(y,x)$ defined  by (\ref{LE1}) is a continuous in
any interior point of the domain $D_{y_0}$. Moreover, for any
$x_0\in \mathbf{R}^1,\ 0<y<y_0$ the following inequality is hold}
\begin{align}
{\rm ess}\sup_{x\leq x_0}|u(x,y)|\leq C(\varepsilon, y_0, x_0,
\|u_0\|), N_{3+\varepsilon}(u_0))\cdot y^{-\frac{1}{4n}}.
\label{Th6_1}
\end{align}

\textbf{Proof. } According to estimates (\ref{GFEst}) and
Cauchy-Bunyakowsky   inequality we have
\begin{align}
\left|\int\limits_{-\infty}^x G(y, x-\xi)u_0(\xi)d\xi\right|\leq C
\int\limits_{-\infty}^x \frac{x-\xi
+1}{y^{\frac{1}{4n}}}|u_0(\xi)|d\xi \leq\nonumber\\
\leq\frac{C(\varepsilon)}{y^{\frac{1}{4n}}}\left(\int\limits_{-\infty}^{x_0}
u_0^2(\xi)(x_0-\xi+1)^{3+\varepsilon}d\xi\right)^\frac{1}{2}\leq\frac{C(\varepsilon,
x_0, \|u_0\|, N_\alpha(u_0))}{y^{\frac{1}{4n}}}.\label{Th6_2}
\end{align}

\begin{align}
\left|\int\limits_x^\infty G(y, x-\xi)u_0(\xi)d\xi\right|\leq
\int\limits_x^\infty
|u_0(\xi)|\frac{1}{y^{\frac{1}{2n+1}}}\exp\left(-c_0\frac{(\xi -x
)^{\frac{2n+1}{2n}}}{y^{\frac{1}{2n}}}\right)d\xi\leq\nonumber\\
\leq\frac{c}{y^{\frac{1}{2n+1}}}\left(\int\limits_x^\infty
u_0^2(\xi)d\xi\right)^{\frac{1}{2}}
\cdot\left(\int\limits_x^\infty\exp\left[-2c_0\frac{(\xi -x
)^{\frac{2n+1}{2n}}}{y^{\frac{1}{2n}}}
\right]d\xi\right)^{\frac{1}{2}}\leq\frac{C\|u_0\|}{y^{\frac{1}{4n+2}}}.
\label{Th6_3}
\end{align}

The inequalities (\ref{Th6_2}), (\ref{Th6_3}) implies the estimate
(\ref{Th6_1}). It is not difficult to see that the condition of
theorem are sufficient for uniform convergence of the integral in
(\ref{LE1}). Hence $u(y,x)$ is continuous.

\subsection{Existence of weak solution for non linear equation}
\label{2B}

In this paragraph we investigate existence of weak solution in the
sense of definition 1.

First we give estimates for the case when initial condition is in
$S(\mathbf{R}^1)$.

\textbf{Lemma 1.} Let $(-1)^{n+k}b_{2k}>0, \ k=\overline{0, n-1}$.
Then the following estimates are hold
\begin{align}
\sup_{0<y<y_0}\|u(y,x\|\leq \|u_0(x)\|, \label{L1_1} \end{align}

\begin{align}
\|u\|_{L_2(0,y_0; H^{n-1}(\mathbf{R}^1))}\leq C\|u_0\|,
\label{L1_2}
\end{align}

\begin{align}
\int\limits_0^{y_0}\int\limits_{x_0-1}^{x_0}\left(\partial_x^nu(y,x)\right)^2dxdy\leq
C(y_0, \|u_0\|), \label{L1_3}
\end{align}

for all $x_0\in \mathbf{R}^1$. Furthermore, if
$N_\alpha(u_0)\leq<\infty$ for some $\alpha\geq 0$ then
\begin{align}
\sup_{0<y<y_0} N_{\alpha}(u(y,x))\leq C(\alpha, y_0, \|u_0\|,
N_\alpha (u_0)), \label{l1_4}
\end{align}

\begin{align}
\int\limits_0^{y_0}N_{\alpha}\left(\partial_x^k
u(y,x)\right)dy\leq C(\alpha, y_0, \|u_0\|, N_\alpha(u_0)), \
k=\overline{1, n-1}, \label{L1_5}
\end{align}

and if $\alpha >0$ then
\begin{align}
\int\limits_0^{y_0}(x_0-x+1)^{\alpha -1}\left(\partial_x^n
u(y,x)\right)^2dxdy\leq C(\alpha, y_0, x_0, \|u_0\|,
N_\alpha(u_0)). \label{L1_6}
\end{align}

\textbf{Proof.} The  estimates (\ref{L1_1}) and (\ref{L1_2}) was
proven in the previous section. To show other estimates we
multiply both sides of equation (\ref{NLEq2}) by
$u(y,x)\cdot\psi_\alpha(x_0-x)$ and integrate in $D_{y}$
\begin{align}
\frac{2n+1}{2}\intly\intlx\left(\partial_x^nu(y,x)\right)^2\psi_\alpha^{\prime}(x_0-x)dxd\eta
+\nonumber\\
+\sum_{k=0}^{n-1}\intly\intlx\left(\partial_x^ku\right)^2\psi_\alpha(x_0-x)dxd\eta+\frac{1}{2}\intlx
u^2\psi_\alpha(x_0-x)dxd\eta\nonumber\\
\leq\sum\intly\intlx\left(\partial_x^k
u\right)^2\sum_{j=1}^{2n+1-2k}A_{kj}\left|\psi_\alpha^{(j)}(x_0-x)\right|dxd\eta+\nonumber\\
+\frac{1}{2}\intlx u_0^2(x)\psi_\alpha(x_0-x)dx
+|\gamma|\intly\intlx |u|^3 \psi_\alpha^\prime(x-0-x)dxd\eta.
 \label{L1P1} \end{align}

Initially we estimate third integral on the right hand side of
(\ref{L1P1})

$$\intlx |u(y,x|^3\psi_\alpha^\prime (x_0-x)dx\leq \sup_x|u(y,x)\sqrt{\psi_\alpha^\prime(x_0-x)}|\intlx u^2(y,x)\sqrt{\psi_\alpha^\prime(x_0-x)}dx.$$

Using inequality (\ref{v'}) we estimate supreme

$$\sup_x |u(y,x)\sqrt{\psi_\alpha^\prime(x_0-x)}|\leq \sqrt 2 \left(\intlx\left|u\sqrt{\psi_\alpha^\prime}\right|
\cdot \left|\partial_x u\sqrt{\psi_\alpha^\prime}+\frac{u\cdot\psi_\alpha^{\prime\prime}}{\sqrt{\psi_\alpha^\prime}}
\right|\right)^{\frac{1}{2}}\leq$$

$$\leq\frac{1}{\gamma}\left(\intlx u^2\cdot |\psi_\alpha^{\prime\prime}|dx\right)^{\frac{1}{2}}+\sqrt 2\gamma\left(\intlx u^2\psi_\alpha^\prime dx\right)^{\frac{1}{4}}
\left(\intlx(\partial_xu)^2\psi_\alpha^\prime dx\right)^{\frac{1}{4}}.$$

Continuing inequality (\ref{L1P1}) we get

\begin{align}
\frac{2n+1}{2}\intly\intlx\left(\partial_x^nu(y,x)\right)^2\psi_\alpha^{\prime}(x_0-x)dxd\eta
+\nonumber\\
+\sum_{k=0}^{n-1}\intly\intlx\left(\partial_x^ku\right)^2\psi_\alpha(x_0-x)dxd\eta+\frac{1}{2}\intlx
u^2\psi_\alpha(x_0-x)dxd\eta\nonumber\\
\leq\sum\intly\intlx\left(\partial_x^k
u\right)^2\sum_{j=1}^{2n+1-2k}A_{kj}\left|\psi_\alpha^{(j)}(x_0-x)\right|dxd\eta
+\frac{1}{2}\intlx u_0^2(x)\psi_\alpha(x_0-x)dx
+\nonumber\\
+ \intly\left(\intlx u^2\cdot
|\psi_\alpha^{\prime\prime}|dx\right)^{\frac{1}{2}}dy+\sqrt
2\gamma^2\intly\left(\intlx u^2\psi_\alpha^\prime
dx\right)^{\frac{1}{4}}
\left(\intlx(\partial_xu)^2\psi_\alpha^\prime
dx\right)^{\frac{1}{4}}dy \label{L1P2} \end{align}

Now we will use mathematical induction method. Let
$0\leq\alpha\leq 1$. Then $|\psi_\alpha^{(k)}(x)|\leq C(k)$,
$k\geq 1$. Consequently we have

\begin{align}
\frac{2n-1}{2}\intly\intlx\left(\partial_x^nu(y,x)\right)^2&\psi_\alpha^{\prime}(x_0-x)dxd\eta
+\sum_{k=0}^{n-1}\intly\intlx\left(\partial_x^ku\right)^2\psi_\alpha(x_0-x)dxd\eta+\nonumber\\
&+\frac{1}{2}\intlx u^2\psi_\alpha(x_0-x)dxd\eta\leq C(y_0,
\|u_0\|). \label{IndStep_1L1}
\end{align}

From the last inequality one can easily get inequalities
(\ref{L1_3}) -- (\ref{L1_6}).

Suppose that that the inequalities (\ref{L1_3}) -- (\ref{L1_6})
are hold for $0<\alpha\leq k$ ($k\in \mathbf{N}$).

Let $k<\alpha\leq k+1$. Taking to account the inequalities
$|\psi_\alpha^{(k)}(x)|\leq C(\alpha) (1+\psi_{\alpha -1}(x))$,
$N_{\alpha -1}(v)\leq N_\alpha(v)+\|v\|^2$ we obtain

\begin{align}
\frac{2n+1}{2}\intly\intlx\left(\partial_x^nu(y,x)\right)^2&\psi_\alpha^{\prime}(x_0-x)dxd\eta
+\sum_{k=0}^{n-1}\intly\intlx\left(\partial_x^ku\right)^2\psi_\alpha(x_0-x)dxd\eta+\nonumber\\
+\frac{1}{2}\intlx u^2\psi_\alpha &(x_0-x)dxd\eta\leq C(\alpha,
y_0, x_0, \|u_0\|, N_\alpha(u_0)). \label{IndStep_1L2}
\end{align}
which implies inequalities (\ref{L1_3}) -- (\ref{L1_6}).

\textbf{Theorem 7.}\ {\it
 Let $u_0(x)\in L_2(\mathbf{R}^1)$ and $(-1)^{n+k}b_{2k}>0, \ k=\overline{0,
 n-1}$. Then there exist a weak solution $u(y,x)$ of Cauchy problem in the
 sense of definition 1 which is in
 $$L_\infty (0,y_0; L_2(\mathbf{R}^1))\bigcap L_2 (0,y_0;
 H^{n-1}(\mathbf{R}^1))\bigcap L_2 (0,y_0; H^n(-r, r))
 $$
for any $r>0$. The solution $u(y,x)$ satisfies estimates
(\ref{L1_1}) -- (\ref{L1_6}) in lemma 1. Moreover
\begin{align}
\lim_{y\to 0}\|u(y,x)-u_0(x)\|= 0. \label{T7_1} \end{align} }

\textbf{Proof.} The case $n=1$ is considered in
\cite{Krujkov1983}. The proof of the theorem for $n>1$ is almost
same as in \cite{Krujkov1983}.

 Let $\lambda (x)\in C_0^\infty (\mathbf{R}^1)$,
$\lambda(x)\geq 0$, ${\rm supp}\lambda(x)\subset [-1, 1]$,
$\intlx\lambda (x)dx =1$.

We put $u_{0h}(x):=\psi_0(x+1/h)\psi_0(1/h -x) u_0^h(x)$ where
$$u_0^h(x)=\frac{1}{h}\intlx \lambda\left(\frac{x-z}{h}\right)u(z)dz.$$
It is easy to see that $u_{0h}(x)\in C_0^\infty(\mathbf{R}^1)$.

We consider the Cauchy problem
\begin{align}
Lu_h(y,x)=\partial_x(u_h^2(y,x)),\ \ \ u_h(0,x)=u_{0h}(x).
\label{u_h1}
\end{align}

Put $I_r=(-r, r)$ and $Q_r=(0,y_0)\times I_r$. According to
inequality $\|u_{0h}(x)\|\leq\|u_0(x)\|$ we have
\begin{align}
\sup_{y}\|u_h(y,x)\|\leq \|u_0(x)\|, \label{u_h2} \end{align}

\begin{align}
\int\limits_{Q_r}\left(\partial_x^n u_h(y,x)\right)^2 dxdy\leq
C(y_0, r, \|u_0(x)\|), \label{u_h3}
\end{align}

\begin{align}
\intly\intlx\left(\partial_x^ku_h(y,x)\right)^2dxdy\leq
C\|u_0\|^2, \ \ k=\overline{1, n-1}. \label{u_h4}
\end{align}

\begin{align}
\int\limits_0^{y_0}\left(\sup_{x\in I_r}|u_h(y,x)|\right)^4dy\leq
C(y_0, \|u_0\|)\left[1+\int\limits_{Q_r}(\partial_x
u_h(y,x))^2dxdy\right]\leq C(y_0, r, \|u_0\|). \label{u_h5}
\end{align}

For $n>1$ the constant in the last inequality does not depends on
$r$.

Using the last estimates and Holder's inequality we have
\begin{align}
&\int\limits_0^{y_0}\left(\int\limits_{-r}^r u_h^2(y,x)(\partial_x
u_h(y,x)dx\right)^{\frac{2}{3}}dy\leq \nonumber\\
\left(\int\limits_0^{y_0}\sup_{x\in I_r}
|u_h(y,x)|^4dy\right)^{\frac{1}{3}}&\cdot
\left(\int\limits_0^{y_0}\int\limits_{-r}^r (\partial_x
u_h(y,x))^2dxdy\right)^{\frac{2}{3}}\leq C(y_0, r,
\|u_0\|).\label{u_h6}
\end{align}

The inequality (\ref{u_h6}) means that $u_h\cdot\partial_x u_h$ is
bounded in $L_{\frac{4}{3}}(0,y_0; L_2(I_r))$. Consequently we
conclude that the set of functions $\{ u_h(y,x), \ h>0\}$ is
bounded in
$$W_r:= \{v(y,x): v\in L_2(0, y_0; H^n(I_r)),\ \partial_y v\in L_{\frac{4}{3}}(0,y_0; H^{-n-1}(I_r))\} .$$

According to the theorem 5.1 in the first chapter of \cite{Lions}
the set of functions $W_r$ is compact in $L_2(0,y_0;
H^{n-1}(I_r))$. So we can select the sequence $h_m, \
\lim_{m\to\infty}h_m =0$ such that $u_{h_m}\to u $ *-weakly in
$L_\infty (0, y_0; L_2(\mathbf{R}^1))$, weakly in $L_2(0, y_0;
H^{n-1}(\mathbf{R}^1))$ and for any $r>0$ :\  $u_{h_m}\to u $ in
 in $L_2(0, y_0;
H^{n-1}(-r,r))$, weakly in $L_2(0, y_0; H^n(-r, r))$.

Using the convergence properties given above one can easily show
that the function $u(y,x)$ satisfies the integral identity
(\ref{IntIdent})and the estimates (\ref{L1_1})--(\ref{L1_6}).

Now we should consider fulfillment of initial condition. For any
positive integer $p$ we define the set $E_p\subset (0,y_0), \ {\rm
mes} E_p=0$ such that for $y\in (0,y_0)\setminus E_p$ the
following relations are hold
$$\|u(y,x)\|\leq \|u_0(x)\|, \ \ \int\limits_{-p}^p \left| u_{h_m}(y,x)-u(y,x)\right|^2dx\to 0.$$
We put $E=\bigcup\limits_{p=1}^{\infty} E_p$, ${\rm mes} E_p =0$.

 Let $\omega (x)\in C_0^\infty (\mathbf{R}^1)$. There exist
positive integer $p$ that ${\rm supp} \omega (x)\in (-p, p)$. We
have
$$\left|\intlx (u_{h_m}(y,x)- u_{0,h_m}(x))\omega(x)dx\right|=
\left|\intly\intlx \partial_y \left( u_{h_m}(\eta
,x)\right)\omega(x)dxd\eta\right|\leq$$

$$\leq\|\omega(x)\|_{H^{n+1}(-p, p)}\cdot \intly \left\|\partial_y u_{h_m}(\eta ,x)\right\|_{H^{-n-1}(-p, p)}d\eta\leq$$

$$\leq\|\omega(x)\|_{H^{n+1}(-p, p)}\cdot\left[\left\|\partial_y u_{h_m}(y,x)\right\|_{L_{4/3}(0,y_0; H^{-n-1}(-p, p))}\right]^{\frac{3}{4}}\cdot y^{\frac{1}{4}}\leq$$

$$\leq C(y_0, p, \|u_0\|)\|\omega(x)\|_{H^{n+1}(-p,p)}\cdot y^{\frac{1}{4}}.$$

Consequently taking limits $m\to\infty$, $y\to 0$ we get
(\ref{IntIC}).

Now we should show (\ref{T7_1}).
$$\|u(y,x)-u-0(x)\|^2\leq 2\|u_0\|^2-2\intlx u(y,x)\cdot u_0(x)dx=$$

$$+2\intlx (u_0(x)-u(y,x))u_{0h}(x)dx+  2\intlx (u_0(x)-u(y,x))(u_0(x) -u_{0h}(x))dx\leq$$

$$+2\intlx (u_0(x)-u(y,x))u_{0h}(x)dx+4\|u_0(x)\|\cdot \|u_0(x)-u_{0h}(x)\|.$$
The last inequality proves (\ref{T7_1}).

The theorem 7 is proved.

\subsection{Continuous Dependence of The Weak Solution on Initial
Data.}\label{2C}

For any $\alpha >0$ any $r>0$ we put

$$\rho_\alpha (x)=\left\{\begin{array}{cl} e^{-x},& x>0,\\
(1-x)^\alpha, & x\leq 0, \end{array}\right.
$$

$$\rho_{\alpha ,r} (x)=\left\{\begin{array}{cl} \rho_\alpha (x),& x \geq -r,\\
(1+r)^\alpha, & x< -r.\end{array}\right.
$$

$$\gamma_{\alpha ,r}(y, \xi ,\eta)=\intlx (\partial_x(y-\eta, x-\xi))^2\rho_{\alpha, r}(x)dx.$$

First we give the following lemmas without proof.

\textbf{Lemma 2.} {\it Let $u(y,x)$ be a weak solution of the
Cauchy problem (\ref{NLEq2}), (\ref{IC2}) and for some
$\varepsilon >0$
$${\rm ess}\sup_y \left[\|u(y,x\|^2+N_{3+\varepsilon}(u(y,x))\right]\leq \infty.$$

Then the following identity is hold a.e. in $D_{y_0}$
\begin{align}
u(y,x)=\intlx G(y, x-\xi)u_0(\xi)d\xi+ (-1)^n\gamma\intly\intlx
u^2(y-\eta, x-\xi)d\xi d\eta. \label{ObrLin}
\end{align}}

\textbf{Lemma 3.} {\it Let $0\leq\eta\leq y\leq y_0$. Then the
following estimates are hold

\begin{align}
\gamma_{\alpha ,r} (y, \xi ,\eta )\leq C(\alpha ,y_0)\rho_{\alpha
,r}(\xi ) (y-\eta)^{-\frac{4}{2n+1}} \ \ \mbox{for  }\ \xi >0,
\label{gamma_r}
\end{align}

\begin{align}
\gamma_{\alpha ,r} (y, \xi ,\eta )\leq C(\alpha ,y_0)\rho_{\alpha
,r}(\xi ) (y-\eta)^{-\frac{4}{2n+1}} (-\xi)^{3+\frac{1}{n}} \ \
\mbox{for }\ \xi <0. \label{gamma_r2} \end{align}}

The proof of lemmas 2 and 3 are analogous as similar lemmas in
\cite{Krujkov1983}.

We define the class of functions
$$\mathbf{K}:=\{u(y,x): M[u,y_0]\leq\infty\},$$
where
$$M[u,y_0]:={\rm ess}\sup_y\left[\|u(y,x)\|^2+ N_{3+\frac{1}{n}}(u(y,x))\right]. $$

\textbf{Theorem 8.} Let the functions $u(y,x)\in \mathbf{K}$ and
$v(y,x)\in\mathbf{K}$ be weak solutions of the Cauchy problem with
initial conditions $u(0,x)=u_0(x)$ and $v(0,x)=v_0(x)$ and for
some $\alpha_0\geq 0$ the quantities $N_{\alpha_0}(u_0(x))$ and
$N_{\alpha_0}(v_0(x))$ are bounded. Then for any
$0<\alpha\leq\alpha_0$
\begin{align}
{\rm
ess}\sup_y\intlx\rho_\alpha(x)\left(u(y,x)-v(y,x)\right)^2dx\leq
C(\alpha, y_0, M[u, y_0], M[v,y_0])\left[\|u_0(x)-v_0(x)\|^2+
N_\alpha (u_0(x)-v_0(x))\right], \label{T8_1}
\end{align}
and if $\alpha=0$ then
\begin{align}
{\rm
ess}\sup_y\intlx\rho_\alpha(x)\left(u(y,x)-v(y,x)\right)^2dx\leq
C(\alpha, y_0, M[u, y_0], M[v,y_0])\cdot\|u_0(x)-v_0(x)\|^2.
\label{T8_2}
\end{align}

\textbf{Proof.} Let $\omega(y,x)$ be a solution of Cauchy problem
for the equation $L\omega(y,x)=0$ with initial condition $\omega
(0,x)=u_0(x)-v_0(x)$ and $W(y,x)=u(y,x)-v(y,x)$. According to
lemma 2 we have
\begin{align}
W(y,x)=\omega(y,x)+(-1)^n\gamma\intly\intlx
[u(\eta,\xi)+v(\eta,\xi)]W(\eta,\xi)\partial_x G(y-\eta,
x-\xi)d\xi d\eta. \label{PT_8_1}
\end{align}

Using the inequality $(a+b)^2\leq 2(a^2+b^2)$ from (\ref{PT_8_1})
we have
\begin{align}
&\intlx W^2(y,x)\rho_{\alpha,r}(x)dx\leq 2\intlx
\omega^2(y,x)\rho_{\alpha,r}(x)dx +\nonumber\\
+2\gamma^2\intlx\rho_{\alpha,r}(x)dx&\left(\intly\intlx(\partial_xG(y-\eta,
x-\xi))W(\eta, xi)[u(\eta,\xi)+v(\eta, \xi)]d\xi
d\eta\right)^2dx=:2I_1+2\gamma^2 I_2.
 \label{PT_8_2}
\end{align}

According to the results from previous section we get
\begin{align}
I_1=\intlx \omega^2(y,x)\rho_{\alpha,r}(x)dx\leq
C(\alpha)\left[\|\omega\|^2+N_\alpha (\omega)\right]\leq C(\alpha,
y_0)\left[\|u_0-v_0\|^2+N_\alpha(u_0-v_0)\right]. \label{PT_8_3}
\end{align}

To estimate the integral $I_2$ we will use Cauchy-Bunyakowski
inequality.

$$
I_2\leq C(\delta, y_0) \intlx\rho_{\alpha,r}(x)\intly
(y-\eta)^\delta \nu(\eta)\intlx (\partial_x G(y-\eta,
x-\xi))\rho^{-1}_{\alpha,r}(\xi)[u(\eta, \xi)+v(\eta, \xi)]^2d\xi
d\eta dx, $$

where $\frac{1}{3}<\delta<1$, $\nu(\eta)=\intlx W^2(\eta,
\xi)\rho_{\alpha,r}(\xi)d\xi$.

Changing integration order  and the Lemma 3 we have

\begin{align}
I_2\leq C(\delta, y_0)\intly (y-\eta)^\delta\nu (\eta)d\eta\intlx
[u(\eta, \xi)+v(\eta, \xi)]^2\rho_{\alpha,
r}(\xi)\gamma_{\alpha,r}(y, \xi,\eta)d\xi\nonumber\\
\leq 2C(\delta, y_0)\intly (y-\eta)^\delta\nu
(\eta)\left[\int\limits_0^\infty [u^2(\eta, \xi)+v^2(\eta,
\xi)]d\xi+\int\limits_{-\infty}^0 (-\xi)^{3+\frac{1}{n}}[u^2(\eta,
\xi)+v^2(\eta, \xi)]d\xi\right]d\eta.
 \label{PT_8_4}
\end{align}

Summarizing inequalities (\ref{PT_8_2}) -- (\ref{PT_8_4}) we
obtain

$$\nu(y)\leq C(\alpha, y_0)\left[\|W\|^2+N_\alpha(W))\right]+
C(\alpha, y_0, M[u,y_0],
M[v,y_0])\intly(y-\eta)^{\delta-\frac{4}{2n+1}}\nu(\eta)d\eta.$$

Solving the last inequality we have $${\rm
ess}\sup_y\intlx\rho_{\alpha, r}(x)W(y,x)dx\leq C(\alpha, y_0,
M[u,y_0], M[v,
y_0])\left[\|u_0(x)-v_0(x)\|^2+N_\alpha(u_0(x)-v_0(x))\right].$$

Right hand side of the inequality does not depends on $r$.
Therefore we can take limit $r\to 0$ and obtain the inequalities
(\ref{T8_1}) and (\ref{T8_2}) for $\alpha =0$. The theorem 8 is
proved.

Thus in this  paper we studied  Cauchy problem for a high order
generalization of KdV equation. In particular, we proved
solvability of the problem for the case of initial function in
S(R1). In addition, using this result the existence of a weak
solution in the case of initial function in L2(R1) and its
continuous dependence on the initial conditions  are shown.

We used Green function, Fourier transform, iteration, averaging of
function, a-priory estimates are used to obtain the above results.
Finally, it follows from the theorems 7 and 8  that if $M[u_0; y]
= M0 < +\infty$, then the Cauchy problem has a solution in the
class K and in this class the solution is unique.


\begin{thebibliography}{88}
\bibitem{Gardner1967} Gardner C.S. et al.(1967) Method for solving KdV
equation. Phys. Rev. Letters, v.19, 1095.
\bibitem{Miura1968} Miura R.M., Gardner C.S., Kruskal M.D. (1968) KdV
equation and generalizations: II Existence of conservation laws
and constants of motions. J. Math. Phys., v.9, 1204.
\bibitem{Djur1991} T.D.Djuraev, S.Abdinazarov. (1991) Dokladi of
Acad. of Sci. USSR, v.320, 6, p. 1305-1309.
\bibitem{Lions} Lions J.-L., Moskva, Mir, 1972.
\bibitem{Shabat1973} Shabat, A.B.(1973) On KdV equation. Doklady Acad.
Sci. USSR, 211, 6, 1310.
\bibitem{Yakupov1975} Yakupov V.M. (1975) On Cauchy problem for
KdV equation. J. Diff. Equ., 11, 3, 556.
\bibitem{Krujkov1983} Krujkov S.N., Faminskii A.V. (1983) Matematicheskiy
Sbornik, 120(162), 3, 396-425.
\bibitem{Gelfand1953}  Gelfand I.M., Shilov G.E. (1953) Uspekhi Math. Nauk USSR,3-54.
\bibitem{Gelfand1955} Gelfand I.M., Shilov G.E. (1955) Docladi
Acad. Nauk USSR, 102:6, 1065-1068.
\bibitem{Gelfand1958} Gelfand I.M., Shilov G.E. Generalized
functions. Some problems of the theory of differential equations.
Moskow, FizMatGIz, 1958.
\bibitem{AbdSob1}S.Abdinazarov, Z.A.Sobirov. Cauchy problem for a nonlinear, high odd
 order equation with multiple characteristics. Proc. of Int. Conf. "Spec. Theory of Diff. Operators and Related Problems ".
  Sterlitamak, Russia, 2003. p. 71.
\bibitem{AbdSob2} S.Abdinazarov, Z.A.Sobirov. On continuous dependent of
generalized solution of Cauchy problem from initial data for high
odd order nonlinear equation. Proc. Int. Russian-Uzbek symposium.
Nalchik,  2003, p.10.
\bibitem{AbdSob3}S.Abdinazarov, Z.A.Sobirov. Cauchy problem for high odd
order equation on   space. Proc. of Int. Conf. "PDE and related
problems of analyses and informatics". Tashkent, 2004 . vol. I. p.
145.
\end{thebibliography}
\end{document}